# High quality factor polariton resonators using van der Waals materials

Michele Tamagnone[1]*, Kundan Chaudhary[1], Christina M. Spaegele[1], Alex Zhu[1], Maryna Meretska[1], Jiahan Li[2], James H. Edgar[2], Antonio Ambrosio[3,4], Federico Capasso[1]*.

[1] Harvard John A. Paulson School of Engineering and Applied Sciences, Harvard University, Cambridge, Massachusetts 02138, USA.

[2] Department of Chemical Engineering, Kansas State University, Manhattan, Kansas 66506, USA.

[3] Center for Nanoscale Systems, Harvard University, Cambridge, MA 02138, USA.

[4] CNST – Fondazione Istituto Italiano di Tecnologia, Via Giovanni Pascoli, Milan, Italy.

*Correspondence to: mtamagnone@seas.harvard.edu, capasso@seas.harvard.edu.

**Abstract**

**We present high quality factor optical nanoresonators operating in the mid-IR to far-IR based on phonon polaritons in van der Waals materials. The nanoresonators are disks patterned from isotopically pure hexagonal boron nitride (isotopes $^{10}$B and $^{11}$B) and α-molybdenum trioxide. We experimentally achieved quality factors of nearly 400, the highest ever observed in nano-resonators at these wavelengths. The excited modes are deeply subwavelength, and the resonators are 10 to 30 times smaller than the exciting wavelength. These results are very promising for the realization of nano-photonics devices such as optical bio-sensors and miniature optical components such as polarizers and filters.**

**Introduction**

One of the most important challenges of nanophotonics is the realization of optical systems that can overcome the diffraction limit. To this aim, the use of simple dielectric materials is not sufficient and alternative materials have to be considered. At visible wavelengths, plasmon polaritons in noble metals have been extensively studied and used to realize e.g. highly subwavelength nanoresonators. More recently, the study of phonon polaritons in SiC and van der Waals (vdW) materials has provided a valid counterpart of visible plasmonics at infrared frequencies [1-11]. On thin flakes (few nm to few hundreds of nm) of vdW materials, phonon polaritons form highly confined guided modes that can propagate across the flakes with high effective index, having propagation velocities tens or hundreds of times smaller than the speed of light. If a vdW flake is patterned into nanostructures, the propagating waves forms standing wave resonances with a discrete spectrum and with relatively high quality factors in the order of hundreds. Unlike other types of optical resonators, this is achieved maintaining deeply subwavelength geometries [7, 9]. Therefore, these resonators are characterized by very large Purcell factors and field enhancement and have attracted significant interest in the nanophotonic community.

In this contribution we will focus on two vdW materials: hexagonal boron nitride (hBN) and α-molybdenum trioxide (αMoO₃). The isotopic purity of the material plays a major role in the propagation length of the phonon polaritons (and hence in the quality factor of the resonators). Natural-abundant nitrogen can be considered already isotopically pure ($^{14}$N constitutes the 99.6% of natural nitrogen), while natural-abundant boron is formed by a mixture of two isotopes – $^{10}$B and $^{11}$B with 19.9% and 80.1% abundance respectively. Isotopically pure hBN can be synthesized from either of isotopes, and both h$^{10}$BN and h$^{11}$BN have demonstrated polariton propagation lengths about three times better than their natural abundance counterpart [6].

The resonant phonon polariton response of both hBN and αMoO₃ can be modelled using with a 3x3 diagonal relative permittivity tensor, with each entry on the diagonal ($\varepsilon_{xx}, \varepsilon_{yy}, \varepsilon_{zz}$) taking the form:

$$\varepsilon = \varepsilon_\infty \left(1 - \frac{(\omega_{LO})^2 - (\omega_{TO})^2}{\omega^2 - i\omega\Gamma - (\omega_{TO})^2}\right) \quad (1)$$

where $\omega = 2\pi f = 2\pi c/\lambda$ is the angular frequency, $c$ is the speed of light, $\lambda$ is the wavelength, $\varepsilon_\infty$ is the high frequency limit of the permittivity, $\Gamma$ is the resonance linewidth, $\omega_{TO}$ and $\omega_{LO}$ are the lower and upper limits of the restrahlen band respectively (i.e. the band where the real part of the permittivity takes negative values, here indicated with RS1, RS2, ...). Equation 1 is equivalent to a Lorentz model of the form:

$$\varepsilon = \varepsilon_\infty + \frac{\varepsilon_{Lorentz}(\omega_{TO})^2}{(\omega_{TO})^2 + i\omega\Gamma - \omega^2} \quad (2)$$

where the resonant frequency corresponds to $\omega_{TO}$ and the Lorentz permittivity $\varepsilon_{Lorentz}$ is given by:

$$\varepsilon_{Lorentz} = \varepsilon_\infty \left(\left(\frac{\omega_{LO}}{\omega_{TO}}\right)^2 - 1\right) \quad (3)$$

hBN is a uniaxial crystal, with in-plane isotropic behavior ($\varepsilon_{xx} = \varepsilon_{yy} \neq \varepsilon_{zz}$), while αMoO₃ is a biaxial crystal ($\varepsilon_{xx} = \varepsilon_{yy} = \varepsilon_{zz}$).

The model parameters for the materials considered in this paper are given in Table 1, and the permittivities are plotted in Figure 1a, b.

**Table 1** - Modelling parameters for hBN and αMoO₃

| Material | Ref. | RS band | Axis | $\omega_{TO}$ (cm⁻¹) | $\omega_{LO}$ (cm⁻¹) | $\Gamma$ (cm⁻¹) | $\varepsilon_\infty$ |
|---|---|---|---|---|---|---|---|
| hBN nat. ab. | [5,6] | RS1 | z | 760 | 825 | 3 | 2.95 |
| | | RS2 | x, y | 1366 | 1610 | 7 | 4.87 |
| h$^{10}$BN | [6] | RS1 | z | 785 | 845 | 1 | 2.5 |
| | | RS2 | x, y | 1394.5 | 1650 | 1.8 | 5.1 |
| h$^{11}$BN | [6] | RS1 | z | 755 | 814 | 1 | 3.15 |
| | | RS2 | x, y | 1360 | 1609 | 2.1 | 5.32 |
| αMoO₃ | [10] | RS1 | y | 545 | 851 | 4 | 5.2* |
| | | RS2 | x | 818 | 974 | 4 | 4.0 |
| | | RS3 | z | 962 | 1010 | 2 | 2.4 |

*A better fit for the αMoO₃ resonators shown in this work is obtained for $\varepsilon_{\infty,yy} = 4.1$

In a resonator, oscillations are caused by the periodic transformation of the stored energy from one form to another. For standalone optical nanoresonators this can be achieved typically in two different ways:

I. Resonators where the stored energy oscillates between quasi-static electric fields and kinetic energy associated to polariton excitation,
II. Resonators where the stored energy oscillates between electric fields and magnetic fields.

Type I resonators (e.g. [7, 9]) are proper polaritonic resonators in the sense that the optical energy is periodically stored in the kinetic energy of the carriers (for plasmons) or the ion cores (for phonons). They rely on the negative real permittivity due to the polariton and can be miniaturized of several order of magnitudes with respect to the exciting wavelength. In contrast, in type II resonators, polaritons are only instrumental to create a high refractive index constant between the nanostructure and the environment, and the resonance is determined by the geometry of the electromagnetic fields supported by the structure. Therefore, these structures have size in the order of the wavelength of used light and high Q-factor resonances can be achieved only using weakly coupled Fano resonances, that can be achieved with complex finely tuned structures [12]. As explained in more details in the Supplementary Information, only type I resonators can achieve deep subwavelength miniaturization, but the quality factor is bounded by the intrinsic polariton quality factor $Q_P$. Using models of available polaritonic materials, it is possible to compute the intrinsic quality factors for polaritons in several materials (see Table 2 and supplementary Information for details). Importantly, for most van der Waals materials, polaritons are anisotropic and $Q_P$ can take different values along each crystal axis.

**Table 2** - Comparison of measured polariton Q factors for different materials. For van der Waals materials, z is the direction orthogonal to the atomic planes (out-of-plane direction).

| Material | Ref. | Polariton | $Q_P$ along x | $Q_P$ along y | $Q_P$ along z |
|---|---|---|---|---|---|
| hBN nat. ab. | [5,6] | Phonon | 195 | | 253 |
| h$^{10}$BN | [6] | Phonon | 775 | | 785 |
| h$^{11}$BN | [6] | Phonon | 648 | | 755 |
| αMoO$_3$ | [10] | Phonon | 204 | 136 | 481 |
| Monolayer graphene | [13] | Plasmon | 130 @ 60K; 25 @ 300K | | (Not defined) |
| SiC | [4] | Phonon | 167 | | |
| Ag (at $\lambda$ = 6 µm) | [14] | Plasmon | 11 | | |
| Au (at $\lambda$ = 6 µm) | [14] | Plasmon | 4 | | |
| Cu (at $\lambda$ = 6 µm) | [14] | Plasmon | 1.25 | | |
| Al (at $\lambda$ = 6 µm) | [14] | Plasmon | 1 | | |

In this contribution we target the realization of type I ultraconfined resonators at mid and far infrared wavelengths with high Q-factors. We achieve this objective by using high Q polaritons in van der Waals materials. The physics of these modes was explained elsewhere including our previous work on natural abundance hBN resonators [7] and will be extended here to other wavelength ranges using αMoO$_3$ [10, 11] and to higher Q factors with isotopically pure h$^{10}$BN and h$^{11}$BN.

**Results**

The resonators are defined by patterning the flakes with the process detailed in Figure 1c and 1d, using e-beam lithography and reactive ion etching (RIE). Because the resonators are characterized using scattering type scanning near field optical spectroscopy (s-SNOM, Figure 1e), the adhesion of the resonators with the substrate must be ensured. For some of our samples (specifically for the h$^{10}$BN resonators) the adhesion was not sufficient to allow s-SNOM measurements, so the resonators were encapsulated using an additional layer of 30 nm chemical vapor deposited (CVD) SiO$_2$.

While our process can pattern hBN into cylindrical pillars with high aspect ratios [7], on MoO$_3$ crystals the etching is slower, causing a shape similar to a truncated cone (Figures 1 f, g). In addition, even though the e-beam pattern consisted in circles, the MoO$_3$ resonators are faceted in an octagonal shape due to direction selectivity of the RIE etching. For hBN we use a substrate of 300 nm SiO$_2$ grown on silicon, since at the working wavelengths in the RS2 band SiO$_2$ has permittivity very close to 1 [7]. For MoO$_3$ we use a high-resistivity silicon substrate which has very low losses across the whole mid-IR range.

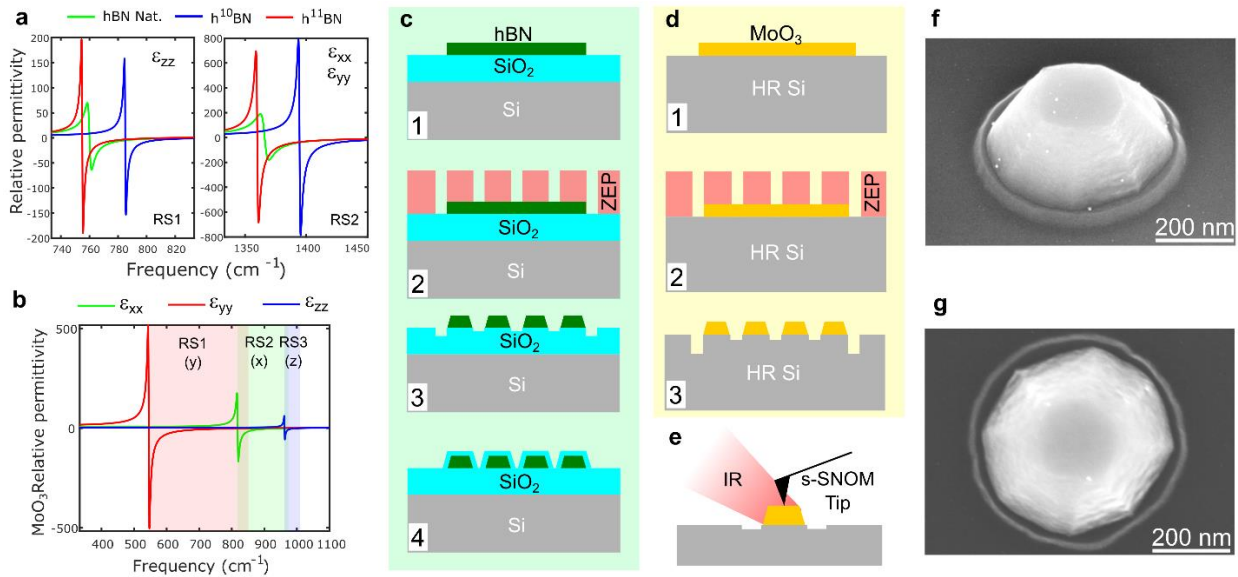

**Figure 1 – Material properties and fabrication of nanoresonators**. **a)** Real part of the permittivity of hBN (natural abundance and isotopically pure). **b)** Real part of the permittivity of αMoO$_3$. **c)** Fabrication of hBN resonators on SiO$_2$/Si substrate. Flakes are exfoliated (1), e-beam lithography with ZEP resist is performed (2), followed by etching and resist strip (3). Optionally, an additional layer of 30 nm of CVD SiO$_2$ is deposited to encapsulate the resonators (4). **d)** Fabrication of αMoO$_3$ resonators on high-resistivity Si, following a similar recipe, without the encapsulation in SiO$_2$. **e)** Measurements with s-SNOM are performed on the samples after fabrication. **f, g)** SEM top and tilted view of a fabricated αMoO$_3$ resonator.

The first considered sample consists of h$^{11}$BN disk of diameter 410 nm and thickness 70 nm without CVD oxide coating. The disk was characterized using a commercial Neaspec scattering type scanning near field optical microscope (s-SNOM) with a nano-FTIR module. In this device the sample is illuminated with an ultrafast mid-IR DFG while a Pt-Ir coated AFM tip operates in tapping mode on it. The tip and the light source are fixed with respect to each other and the sample table is scanned underneath the tip. The light scattered by the tip (modulated at the tip tapping frequency) is then demodulated with an interferometer and a lock-in amplifier, to reconstruct both phase and amplitude of the s-SNOM signal over the optical signal bandwidth. This allows to retrieve the full spectrum of the resonator. We first perform a hyperspectral imaging scanning the sample along its diameter and collecting a spectrum for each point. The amplitude can then be represented in a map as a function of the frequency and the tip position (Figure 2a), which allows to identify the excited modes. The details of the geometry of the resonances in h-BN nanodisks was covered in our previous work [7] where we identified bright dipole modes and dark central modes, among others. In Figure 2a we can clearly see the bright dipole fundamental mode at 1450 cm$^{-1}$ and a central mode at 1520 cm$^{-1}$. Considering here the fundamental dipole resonance, we notice that the resonance remains visible when the tip is tapping outside the nanoresonators, since the fringing fields of the optical mode extends beyond the nanodisk boundary. Importantly, the resonance feature becomes much narrower in this regime than when the tip is directly tapping on the disk. This is due to the fact that the tip perturbs the sample and couples a significant portion of the energy stored in the resonator to the environment, reducing the Q factor and slightly red-shifting the resonance frequency $\omega_0$. Figure 2b shows a single spectrum measured with the tip tapping 100 nm away from the disk, which show a resonance quality factor of Q = 332.

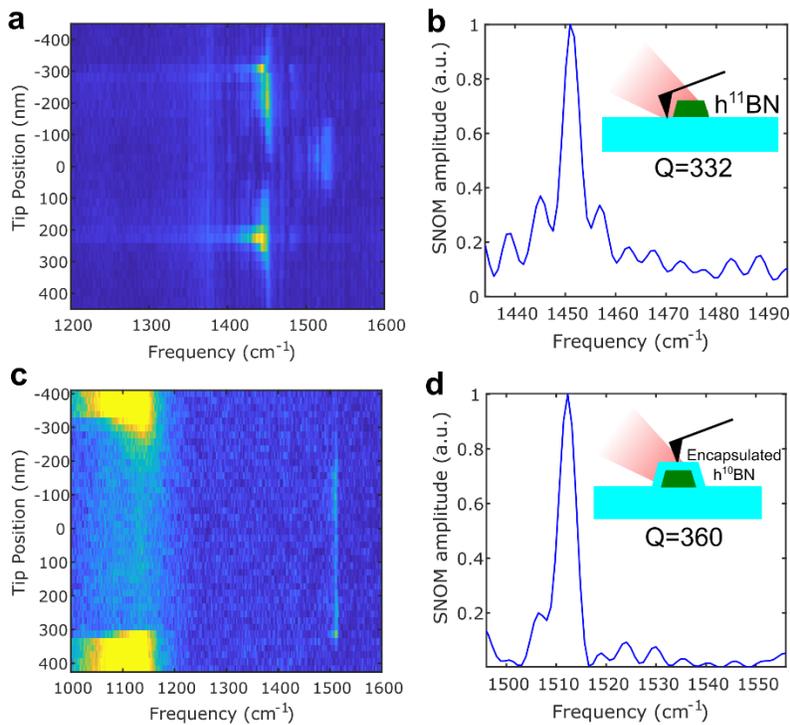

**Figure 2 – s-SNOM characterization of h$^{10}$BN and h$^{11}$BN resonators**. **a)** Hyperspectral s-SNOM scan of a h$^{11}$BN disk of diameter 410 nm and thickness 70 nm (the tip is scanned across the disk and a spectrum is taken for each point). **b)** Single spectrum taken with the s-SNOM tip 100 nm away from the same disk. **c)** Hyperspectral s-SNOM scan of a h$^{10}$BN disk (diameter 630, thickness 80 nm) encapsulated with a 30 nm CVD SiO$_2$ layer. **d)** Single spectrum taken on the same disk. The tip is separated from the hBN by the CVD oxide layer, limiting the tip interaction with the disk and preserving the high quality factor.

The second considered sample is a h$^{10}$BN disk of diameter 630 nm and thickness 80 nm with CVD oxide encapsulation. The encapsulation ensures that the resonator is not detached during the s-SNOM scan and has also an important effect on the resonances: as shown in Figure 2 c, the fundamental resonance is now very narrow even when the tip scans directly on top of the resonator. This is due to the separation between the s-SNOM tip and the hBN by the CVD oxide layer, which limits the interaction of the tip with the sample ensuring a high quality factor of Q = 360.

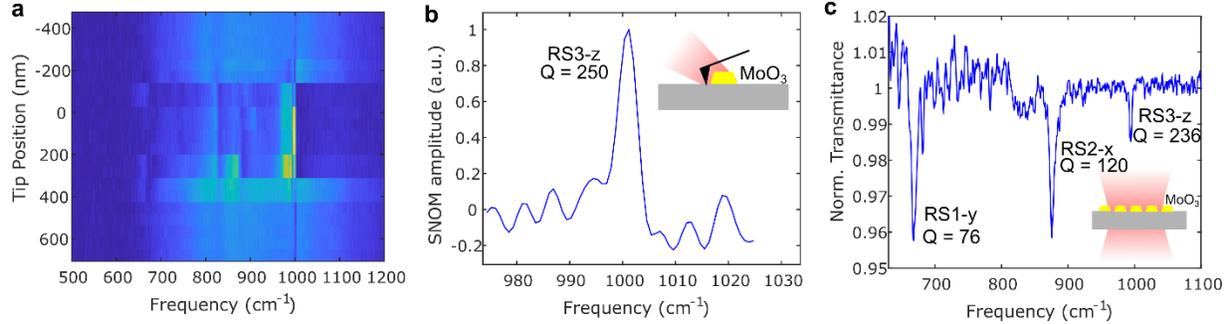

**Figure 3 – Characterization of MoO3 nanoresonators.** The sample consists of an array of MoO3 resonators shaped as conical frusta (see Figure 1f, g) Top diameter is 240 nm, bottom diameter 460 nm and thickness 160 nm. The array is a square lattice with x and y periodicities of 1 μm. **a)** s-SNOM hyperspectral scan. To capture modes in all bands, the tip is scanning at an angle of 45 degrees with respect to the x-y optical axes of MoO3. **b)** Single spectrum (normalized and background-subtracted) acquired with s-SNOM tip 200 nm away from the disk. **c)** Transmission spectrum of the array taken with a far field FTIR microscope.

Finally, we studied an array of $MoO_3$ resonators both via near field and far field techniques (Figure 3). Because $MoO_3$ is a biaxial material, three distinct fundamental dipole resonance are expected for each optical axis. From our near field scans (Figure 3a), only the highest frequency resonance is visible. This resonance is associated to the RS3 band and it is polarized along the z axis. Its Q factor can be estimated, as for the $h^{11}BN$ case, by sampling the field 200 nm away from the nano-article, and it is estimated to be 250. We also performed far field measurement using a far field spectrometer coupled with a mid-IR microscope (Bruker Lumos), and the obtained transmissivity spectrum is shown in Figure 3c. The used beam is unpolarized and can couple well to both the x and y fundamental resonances. Due to imperfections in the array, the z resonance is also visible though very faint, and its quality factor is slightly reduced due to the intrinsic variability in the size of the particles of the array.

**Discsussion**

Table 3 summarizes the measured quality factors of the resonators presented in our work. The highest obtained quality factor is Q = 360 for the encapsulated $h^{10}BN$ resonator. The quality factor of these resonators is bounded by the theoretical limit $Q_{\max}$ (see supplementary Information):

$$Q \leq Q_{\max} = Q_P \frac{\omega_0}{\omega_{TO}} \qquad (4)$$

For all the resonators the quality factor Q is significantly below the theoretical limit. This is due to imperfection in the fabricated shape and to radiative losses in the substrate. Nevertheless, the achieved

Q factor is to the best of our knowledge the highest ever achieved for polaritonic resonators in the mid-IR at room temperature.

**Table 3** – Summary of the resonant frequencies and Q factors of the studied resonators.

| Material | Optical axis | $\omega_{TO}$ (cm$^{-1}$) | $Q_P$ | Meas. type | $\omega_0$ (cm$^{-1}$) | $Q$ | $Q_{max}$ |
|---|---|---|---|---|---|---|---|
| h$^{10}$BN | x, y | 1394.5 | 775 | s-SNOM | 1512 | 360 | 840 |
| h$^{11}$BN | x, y | 1360 | 648 | s-SNOM | 1451 | 332 | 691 |
| αMoO$_3$ | X | 818 | 204 | FTIR microscope | 876 | 120 | 218 |
| αMoO$_3$ | Y | 545 | 136 | FTIR microscope | 669 | 76 | 166 |
| αMoO$_3$ | z | 962 | 481 | s-SNOM | 1001 | 250 | 500 |
| αMoO$_3$ | z | 962 | 481 | FTIR microscope | 994 | 236 | 497 |

**Acknowledgement**

This work was supported by the NSF EFRI, award no. 1542807. This work was performed in part at the Center for Nanoscale Systems (CNS), a member of the National Nanotechnology Coordinated Infrastructure Network (NNCI), which is supported by the National Science Foundation under NSF award no. 1541959. M.T. acknowledges the support of the Swiss National Science Foundation (SNSF) grant no. 168545 and 177836. The hBN crystals growth was supported by the National Science Foundation, award number 1538127.

**References**

[1] D.N. Basov, M. M. Fogler, and F. J. García De Abajo. "Polaritons in van der Waals materials." *Science* **354**, aag1992 (2016).

[2] T. Low et al. "Polaritons in layered two-dimensional materials." *Nature materials* **16**, 182 (2017).

[3] J. D. Caldwell et al. "Low-loss, infrared and terahertz nanophotonics using surface phonon polaritons." *Nanophotonics* **4,** 44 (2015).

[4] P. B. Catrysse and S. Fan. "Near-complete transmission through subwavelength hole arrays in phonon-polaritonic thin films." *Physical Review B* **75**, 075422 (2007).

[5] S. Dai et al. "Tunable phonon polaritons in atomically thin van der Waals crystals of boron nitride." *Science* **343**, 1125 (2014).

[6] A. Giles, et al. "Ultralow-loss polaritons in isotopically pure boron nitride." *Nature materials* **17**, 134 (2018).

[7] M. Tamagnone et al. "Ultra-confined mid-infrared resonant phonon polaritons in van der Waals nanostructures." *Science Advances* **4**, eaat7189 (2018).

[8] A. Ambrosio et al. "Selective excitation and imaging of ultraslow phonon polaritons in thin hexagonal boron nitride crystals." *Light: Science & Applications* **7**, 27 (2018).

[9] M. Autore et al. "Boron nitride nanoresonators for phonon-enhanced molecular vibrational spectroscopy at the strong coupling limit." *Light: Science & Applications* **7**, 17172 (2018).

[10] Z. Zheng, et al. "A mid-infrared biaxial hyperbolic van der Waals crystal." *Science Advances,* In press (2019).

[11] W. Ma et al. "In-plane anisotropic and ultra-low-loss polaritons in a natural van der Waals crystal." *Nature* **562**, 557 (2018).

[12] A. A. Yanik et al. "Seeing protein monolayers with naked eye through plasmonic Fano resonances." *Proceedings of the National Academy of Sciences* **108**, 11784 (2011).

[13] G. X. Ni et al. "Fundamental limits to graphene plasmonics." *Nature* **557**, 530 (2018).

[14] P. Tassin et al. "A comparison of graphene, superconductors and metals as conductors for metamaterials and plasmonics." *Nature Photonics* **6**, 259 (2012).